# Synthesis of Metal-Semiconductor Heterojunctions in Single Graphene Nanoribbons

**Ziyi Wang[1,2,3,6]†, Kaitlin Slicker[4]†, Weichen Tang[1,2]†, Boyu Qie[4,3,5]†, Rafal Zuzak[1], Haowen Pang[1], Yudi Huang[1,7], Xinheng Li[1,8], Peter H. Jacobse[1,9], Steven G. Louie[1,2]*, Felix R. Fischer[4,2,3,5]*, Michael F. Crommie[1,2,3]***

[1]Department of Physics, University of California, Berkeley, CA 94720, USA.

[2]Materials Sciences Division, Lawrence Berkeley National Laboratory, Berkeley, CA 94720, USA.

[3]Kavli Energy NanoSciences Institute at the University of California Berkeley and the Lawrence Berkeley National Laboratory, Berkeley, CA 94720, USA.

[4]Department of Chemistry, University of California, Berkeley, CA 94720, USA.

[5]Bakar Institute of Digital Materials for the Planet, Division of Computing, Data Science, and Society, University of California, Berkeley, CA 94720, USA.

[6]Present address: Department of Chemistry, Columbia University, New York, NY 10027, USA.

[7]Present address: Department of Physics, University of Illinois Urbana-Champaign, Urbana, IL 61801, USA.

[8]Present address: Department of Physics, Cornell University, Ithaca, NY, 14853, USA.

[9]Present address: Department of Materials Science and Nanoengineering, Rice University, Houston, TX 77005, USA.

†These authors contributed equally to this work.

*Corresponding authors.

**Abstract:**

Graphene nanoribbons (GNRs) are a highly tunable class of one-dimensional (1D) quantum materials that can be fabricated through bottom-up synthesis. Precise control over GNR band gaps and band-edge alignments has established them as a promising nanoelectronics platform, but forming high-quality electronic interfaces remains challenging. Here we combine sequential on-surface synthesis and scanning tunneling microscopy (STM) induced dehydrogenation to directly write metallic 7-iGNR segments into otherwise semiconducting $H_2$-7-iGNRs without changing the GNR width or carbon backbone connectivity. The resulting metallic segments exhibit nonzero spectral weight at the Fermi level ($E_F$) and spatially extended electronic states, consistent with two dispersive bands crossing $E_F$. The new metallic states are well-described by an extended Su-Schrieffer-Heeger zigzag-ladder Hamiltonian. Scanning tunnelling spectroscopy (STS) and first-principles calculations show that the valence band edge of adjacent, connected semiconducting GNR segments lies close to $E_F$ for the metallic segments. STS spectra recover bulk GNR properties within ~1 nm on either side of the metal-semiconductor GNR interface, with no discernable depletion region. The observed GNR band edge alignment is consistent with a small energy barrier for hole injection, thus suggesting a possible route towards atomically precise, low-barrier *p*-type contacts for future GNR-based electronic devices.

## Introduction

Engineering the atomic structure of bottom-up-synthesized graphene nanoribbons (GNRs) is a versatile strategy for tuning intrinsic GNR electronic properties since it allows experimental realization of designer one-dimensional (1D) Hamiltonians.[1-6] This has led to the production of ~1 nm width GNRs and GNR heterostructures having a wide range of tunable semiconducting energy gaps.[7-12] The incorporation of GNRs into useful nanotechnological applications, however, has been hampered by difficulty in making electrical contact between semiconducting GNRs and metallic electrodes due to the formation of large Schottky barriers.[13-19] One strategy for overcoming this problem is to create atomically-precise interfaces between metallic and semiconducting GNR segments that allow GNR Schottky barriers to be tuned by integrating bottom-up metallic contacts directly into single GNRs. Some progress has been made in this direction through the linkage of semiconducting GNR segments having different widths and backbone structures,[20-26] as well as STM-based formation of semiconducting heterostructures directly within individual GNRs.[27] Two significant challenges that remain are, first, the creation of GNRs with broadly dispersive *metallic* bands and, second, how to connect them seamlessly to semiconducting GNRs via atomically-precise interface structures. The metallic GNRs made up to now have the disadvantage that they are formed by daisy-chaining topological interface states[28-30] that result in narrow bands that are vulnerable to instabilities from strong interaction effects.[29-31] No strategy has yet been shown that results in controllable, seamless integration of metallic and semiconducting GNR segments from the bottom up. Overcoming these challenges and developing a better understanding of GNR metal-semiconductor interfaces is essential for realizing high-quality all-carbon contacts, interconnects, and logic elements for future nanoelectronics applications.[20-26,28,32]

Here we demonstrate the synthesis of atomically-precise bottom-up GNRs that can be toggled from an initial semiconducting state to a broadly dispersive two-channel metallic state through selective cleavage of hydrogen atoms along its edges using STM tip manipulation. The ability to locally remove hydrogen atoms allows metallic GNR segments to be directly *written* into otherwise semiconducting GNRs, enabling controllable fabrication of bottom-up single-GNR metal-semiconductor heterojunctions that share a common GNR backbone. STS measurements reveal that the characteristic bulk GNR electronic structures of the two segments are recovered over a distance of only ~1 nm across the metal-semiconductor interface. The dehydrogenated metallic state of this uniquely flexible GNR platform exhibits the electronic structure of a nearly ideal quantum zigzag ladder (Figure 1a).[33] In contrast, the presence of $sp^3$-hybridized carbon atoms (methylene groups) along *fully hydrogenated* edges of the GNR disrupts the extended carbon π-electron network, thereby breaking its symmetry and leading to semiconducting GNR segments. This behavior is seen in both our experimental STM measurements as well as in our theoretical DFT-based *ab initio* simulations. Spatially-resolved spectroscopy of substrate-supported metal-semiconductor heterojunctions additionally reveals valence band-edge alignment with the GNR Fermi level ($E_F$) and no resolvable extended depletion region. Such behavior is consistent with a small energy barrier for hole injection, a property that is expected to be advantageous for producing low-resistance electrical contacts in future GNR-based nanodevices.

Our strategy for designing a tunable quantum zigzag-ladder GNR is based on the symmetric Su-Schrieffer-Heeger (SSH) Hamiltonian with one orbital per site and including both nearest-neighbor (NN) ($t_1$) and next-nearest-neighbor (NNN) ($t_2$) electron hopping between orbitals (Figure 1a).[34,35] Achieving broadly dispersive metallic bands requires reduced spacing between the orbitals to increase $t_1$, but this also increases the magnitude of NNN hopping $t_2$. Such behavior is captured in the SSH Hamiltonian as follows:

$$\hat{\mathcal{H}} = \left[\left(-t_1 \sum_i a_i^\dagger b_i - t_1 \sum_i a_i^\dagger b_{i+1} - t_2 \sum_i a_i^\dagger a_{i+1} - t_2 \sum_i b_i^\dagger b_{i+1}\right) + \mathrm{H.c.}\right] + \varepsilon \sum_i \left(a_i^\dagger a_i + b_i^\dagger b_i\right). \quad (1)$$

Here $a_i^\dagger$ ($a_i$) is a creation (annihilation) operator for electrons at site $i$ in the top leg, $b_i^\dagger$ ($b_i$) is the same for the bottom leg, and $\varepsilon$ represents an onsite potential. This leads to a pair of bands that are degenerate at $\boldsymbol{k} = \pi/a$ (where $a$ is the lattice constant shown in Figure 1a), as follows:

$$E(k) = \varepsilon + 2\, t_2 \cos ka \pm 2\, t_1 \cos\left(\frac{ka}{2}\right). \quad (2)$$

For an electron filling of one electron per site (two electrons per unit cell) this results in a metallic ground state.

A scheme for translating this model into the chemical structure of a 7-iGNR is illustrated in Figure 1b. The 7-iGNR is closely related to the well-known 7-AGNR parent class assembled from anthracene building blocks,[1] but exhibits an alternating pattern of laterally fused benzo[*f*]isoindenyl groups along both edges. We can gain insight into the electronic structure of 7-iGNRs through a defect model in which a single 6-membered ring along the edge of a 7-AGNR is replaced by a 5-membered ring (Figure 1c). DFT simulation of this scenario results in two defect states in the 7-AGNR energy gap that are split by $\Delta E \sim 0.4$ eV (labeled *states 1* and *2* in Figure 1d). The wavefunctions for the two defect states are shown in Figure 1c and are localized on the 5-membered ring, but exhibit orthogonal orbital symmetry. At charge neutrality the two defect states accommodate three electrons with *state 2* fully occupied and *state 1* being only half filled (Figure 1d). The electronic structure of the 7-iGNR shown in Figure 1b can be qualitatively thought of as a periodic array of such defects. Here the two defect-derived states hybridize with neighboring cells and each defect state forms a separate quantum zigzag ladder as described by Eq. (1). This hybridization broadens the localized defect levels into two overlapping dispersive bands that cross $E_F$, giving rise to a two-band metallic electronic structure.

## Surface-Assisted Growth of 7-iGNRs

Guided by this strategy, we designed a molecular precursor for the 7-iGNR, 9,9'-dibromo-2,2',3,3'-tetrahydro-1*H*,1'*H*-4,4'-bi(cyclopenta[*b*]naphthalene) (**1**) as shown in Figure 2a (molecular synthesis and characterization are described in the Supplementary Information). Extended 7-iGNRs were prepared following established surface-assisted bottom-up GNR growth protocols. **1** was first sublimed in ultrahigh vacuum (UHV) from a Knudsen cell evaporator onto a clean Au(111) substrate held at $T = 24$ °C. Figure 2b shows a representative topographic STM image of a densely packed monolayer of **1** self-assembled into a square lattice. On-surface step-growth polymerization of molecular precursors was induced by annealing the molecule-decorated surface to $T = 180$ °C for 20 min. STM images of the resulting structures show extended linear chains of *poly*-**1** that laterally assemble into closely packed islands (Figure 2c). A second annealing step at $T = 300$ °C (20 min) was performed to induce thermal cyclodehydrogenation, giving rise to fully fused GNR backbones (Figure 2d). The positions and relative registry of the 5-membered rings are encoded by precursor **1** and the geometry of Ullmann coupling, while cyclodehydrogenation primarily forms the remaining 6-membered rings required to fuse the backbone. Bond-resolved STM (BRSTM) imaging reveals periodic alternation of 5- and 6-membered rings lining the edges of the resulting GNRs (Figure 2d). Throughout the growth irregular 5-ring sequences that result from the occasional flipping of a benzo[*f*]isoindenyl group or incomplete cyclodehydrogenation represent the most common defects in $H_2$-7-iGNRs. In the following discussion we will focus on well-defined defect-free GNR segments whose alternating 5-/6-membered-ring structure was directly verified by BRSTM (as seen in Figure 2d).

Based on the molecular design of **1** and the on-surface reaction sequence shown in Figure 2a, the as-grown GNRs retain $sp^3$ hybridized methylene ($–CH_2–$) groups at the apex of the 5-membered-rings. This is supported by topographic STM images that show the 5-membered rings as apparent depressions (~10 pm) relative to 6-membered rings (Extended Data Fig. 1). We refer to this partially hydrogenated intermediate GNR as $H_2$-7-iGNR. The presence of methylene groups prevents the formation of the radical states associated with the benzo[*f*]isoindenyl groups, effectively gapping out the metallic states expected for 7-iGNRs and confining $\pi$-conjugation to the naphthalene cores along the $H_2$-7-iGNR backbone.

To confirm this intuitive chemical picture we characterized the electronic structure of $H_2$-7-iGNRs using STS. Figure 3a shows a representative d$I$/d$V$ point spectrum acquired above the position of a 5-membered ring on as-grown $H_2$-7-iGNR (d$I$/d$V$ spectra reflect the local density of states (LDOS) under the STM tip[36]). The two most prominent features bracketing $E_F$ are steep rises seen at $V_s = +1.40$ V (for increasing $V_s$) (*feature* 1) and at $V_s = –1.30$ V (for decreasing $V_s$) (*feature* 3). Closer to $E_F$ we observe a third (smaller) characteristic feature, a double peak centered at $V_s = –0.20$ V (*feature* 2). We assign *feature* 1 to the $H_2$-7-iGNR conduction band edge (CBE) and *feature* 2 to the $H_2$-7-iGNR valence band edge (VBE).

From the positions of *features* 1 and 2 we extract an electronic bandgap $E_{g,exp}$ = 1.6 ± 0.3 eV for $H_2$-7-iGNRs on Au(111). This is corroborated by d*I*/d*V* maps of $H_2$-7-iGNR recorded over the range –0.20 V < $V_s$ < +1.40 V that show the featureless LDOS typically seen for semiconducting gaps (Extended Data Fig. 2).[9] d*I*/d*V* maps recorded at biases corresponding to *features* 1, 2, and 3, on the other hand, show distinct nodal patterns. *Feature* 1, for example, shows a faint brick-like pattern (Figure 3b) while *features* 2 and 3 show distinct striped patterns across the backbone of the GNR (Figures 3c,d). The bright protrusions seen in *feature* 3 coincide with the positions of 6-membered rings lining the $H_2$-7-iGNR edges (Figure 3d, Extended Data Fig. 2).

These experimental features are consistent with *ab initio* DFT-LDA simulations of freestanding $H_2$-7-iGNRs (Figures 3e–i). Simulated $H_2$-7-iGNR DOS (Figure 3h) and band structure (Figure 3i) both exhibit a bandgap of 0.8 eV. Taking into account the well-known tendency of DFT-LDA to underestimate energy gaps,[7,8] the experimentally measured gap of 1.6 eV is in reasonable agreement with our DFT simulations. The simulated LDOS maps (Figures 3e–g) also reasonably agree with the experimental d*I*/d*V* maps and further support our assignment of *features* 1 and 2 to the CBE and VBE, respectively.

**Tip-Induced Metallicity**

Inspired by prior studies showing that H-atoms (e.g., methylene groups) lining the edges of nanographenes (NGs) can be cleaved via STM tip manipulation,[37-48] we sought to induce dehydrogenation of $H_2$-7-iGNRs into metallic 7-iGNRs by locally applying voltage pulses. Figure 2e shows the STM *I–V* characteristic recorded while holding an STM tip ~5 Å above the edge of the $H_2$-7-iGNR shown in Figure 2d (at the site of the red cross). As $V_s$ is swept from 0 V to +2.5 V we observe a sudden drop in current at $V_D$ ~ +2.3 V. This drop is irreversible, is independent of polarity, and appears only for voltages above a threshold of $|V_s| \geq 2.0$ V. Similar switching behavior has been observed for other NG systems and is known to occur for tip-induced cleavage of C–H bonds.[40,42,44]

Low-bias BRSTM images recorded in the same region where tip manipulation has been performed show amplitude modulation features commonly associated with the presence of electronic states very close to $E_F$—a hallmark of metallicity in GNRs (Figure 2f).[29,30] Rather than being confined to the cleavage point of a single C–H bond, however, these electronic states extend uniformly across the entire width of the ribbon and ~2 nm along the length from the position of the tip during the pulse. Single voltage pulses thus appear to cleave multiple C–H bonds, giving rise to metallic 7-iGNR segments centered at the position of the applied tip-pulse. Analogous collective behavior has been observed in the decoupling of GNRs from strongly interacting substrates, but is uncommon for tip-induced hydrogen abstraction.[49] Such spatially delocalized hydrogen removal may involve excitation through extended ribbon states followed by vibrational-energy redistribution, or a short sequential cascade in which an initial dehydrogenation event facilitates cleavage at neighboring methylene sites.

We characterized the electronic structure of 7-iGNRs obtained via tip-induced dehydrogenation by recording d*I*/d*V* point spectra on GNR segments subjected to the tip manipulation protocol. Figure 3j shows a representative d*I*/d*V* spectrum for a 7-iGNR obtained after tip-pulsing (our STS measurements focused on 7-iGNR segments of length $L \gtrsim$ 5nm to avoid confinement effects). Three prominent features unique to 7-iGNRs are observed: a broad peak at $V_s$ = +0.80 V (*feature* 4), a narrower peak at $V_s$ = +0.15 V (*feature* 5), and the onset of a shoulder at $V_s$ = –1.20 V (*feature* 6). d*I*/d*V* maps recorded at a bias corresponding to *feature* 4 show intensity patterns reminiscent of the zigzag ladder model of Figure 1a (Figure 3k). As bias is decreased to *feature* 5 the d*I*/d*V* map evolves into a vertically striped phase that is modulated by a longer wavelength along the ribbon axis (Figure 3l). The d*I*/d*V* intensity pattern at *feature* 6 exhibits diamond-shaped structures at the positions of the 5-membered rings (Figure 3m). In contrast to the 1.6 ± 0.3 eV gap observed for $H_2$-7-iGNR, the activated 7-iGNR segments exhibit nonzero d*I*/d*V* intensity down to zero bias, with no resolvable energy gap at $E_F$. The diversity of nodal patterns spanning the range of –1.20 V < $V_s$ < +0.80 V for 7-iGNRs is consistent with an electronic structure in which more than one dispersive band crosses $E_F$ (Extended Data Fig. 3).

*Ab initio* DFT-LDA simulations of 7-iGNRs help us to better understand the origin of metallicity in this GNR. The band structure of a free-standing 7-iGNR is shown in Figure 3r and exhibits two highly dispersive bands crossing $E_F$ (black lines) (charge density wave and magnetic ground states are ruled out at the DFT-LDA level). These bands are degenerate at $E–E_F$ = –0.45 eV and give rise to a pair of van Hove singularities (VHS1 and VHS2) that yield DOS peaks at $E–E_F$ = +0.33 eV and +0.80 eV (Figure 3q). A third VHS at $E–E_F$ = –0.87 eV (VHS3) is seen at the bottom of the lower metallic band due to a gap arising from an avoided crossing. Simulated 7-iGNR LDOS maps at energies corresponding to VHS1, VHS2, and VHS3 (Figures 3n–p) are in reasonably good agreement with the experimental d*I*/d*V* maps for *features* 4, 5, and 6 (Figures 3k–m), supporting the assignment of experimental *features* 4, 5 and 6 to van Hove singularities of the 7-iGNR metallic band manifold. The calculated bandwidth from VHS1 to VHS3 ($\Delta E_{DFT} \approx$ 1.7 eV) also reasonably matches the experimental bandwidth from *feature* 4 to *feature* 6 of 2.0 eV. Simulations of 7-iGNR electronic structure that included the Au(111) substrate were also performed, leading to very similar results.

We tested whether the 7-iGNR electronic structure can be *quantitatively* described as a quantum zigzag ladder by performing a Wannier analysis of the Bloch states for the bands near $E_F$ obtained from our *ab initio* DFT-LDA calculations. Such analysis is useful for determining the most appropriate tight-binding model for a periodic system since it allows the *ab initio* Hamiltonian to be recast into a tight-binding form with no adjustable parameters. To obtain maximally localized Wannier functions we found it necessary to include the four 7-iGNR bands closest to $E_F$. This procedure yields two Wannier functions per 5-membered ring, analogous to the two defect states observed for a pristine 7-AGNR containing a single 5-membered ring defect as shown in Figures 1c,d. By transforming the DFT Hamiltonian into the Wannier basis and truncating small hopping terms we obtain an accurate tight-binding description with only NN and NNN hopping terms, which corresponds precisely to the extended SSH Hamiltonian of Eq. (1) (see Supplementary Information Note S1 for the full DFT Hamiltonian before truncation). Figure 1b shows the two coexisting electronic networks associated with the two Wannier functions (Wannier states 1 and 2), both of which resemble the zigzag ladder of Figure 1a and have the same orbital structure as the *defect states* 1 and 2 in Figure 1c. The red lines in Figure 3r show the tight-binding band structure that results from these two coexisting zigzag Wannier-function networks (Extended Data Table 1 shows the hopping parameters for the truncated model). The two bands crossing $E_F$ thus arise from two zigzag ladder networks and result in a two-band metallic electronic structure for 7-iGNRs. This makes 7-iGNRs a candidate *two-channel* conductor and suggests that extended, low-disorder 7-iGNRs could support two independent quantum conductance channels, analogous to metallic single-walled carbon nanotubes.[50-53]

**Fabricating a Single-GNR Metal–Semiconductor Heterojunction**

Having established that STM tip-induced dehydrogenation converts semiconducting $H_2$-7-iGNRs into metallic 7-iGNRs, we set out to explore the electronic properties of intra-ribbon metal-semiconductor heterojunctions. Figure 4a shows a BRSTM image of a single GNR featuring a 2.5 nm long segment of metallic 7-iGNR obtained by tip-pulsing embedded in an otherwise semiconducting $H_2$-7-iGNR. The detailed evolution of the electronic structure across a 7-iGNR/$H_2$-7-iGNR metal-semiconductor interface can be seen in Figure 4b. Here $dI/dV$ is plotted as a function of $V_s$ and position $x$ across a single interface (the vertical dashed line marks the location of the metal-semiconductor interface). At distances $|x| > 1$ nm on either side of the interface the $dI/dV$ spectra show characteristic features of the pristine $H_2$-7-iGNR and 7-iGNR. The semiconducting VBE is observed to lie close to $E_F$ and no extended depletion region can be resolved for the Au(111)-supported junction.

To gain a deeper understanding of the single-ribbon metal-semiconductor heterojunction behavior observed here, we performed an *ab initio* DFT-LDA simulation of a freestanding 7-iGNR/$H_2$-7-iGNR structure as shown in Figure 4c (aimed at modeling the experiment shown in Figure 4b). Figure 4d shows a plot of the resulting LDOS as a function of position and energy. At distances of $|x| > 1$ nm from the interface the junction recovers the unperturbed electronic structure of pristine 7-iGNR and $H_2$-7-iGNR, showing signatures of van Hove singularities on the metallic side and band edge features on the semiconducting side similar to the experiment. Our finite, freestanding junction simulation shows no extended electrostatic variation, such as the signatures of electrostatic band-bending commonly seen for macroscopic Schottky barriers. The simulation also shows no new states within the semiconducting $H_2$-7-iGNR gap near the interface. It is notable that $E_F$ for the metallic 7-iGNR segment in our simulation aligns very closely with the VBE of the semiconducting $H_2$-7-iGNR segment, similar to experiment, suggesting that the freestanding junction model captures the essential heterojunction electronic structure. The supporting Au(111) substrate of the experiment (which is not included in the simulation) is expected to screen long-range built-in fields and may modify precise band-edge energy positions through charge transfer, hybridization, and image-charge screening. Overall, however, the band-edge energy alignment observed both in experiment and theory for this metal-semiconductor heterojunction is consistent with a small energy barrier for hole injection. This implies that GNR-based devices incorporating bottom-up intra-ribbon contacts as shown here could achieve substantially reduced contact resistance for *p*-type operation compared to previous top-down GNR contacting architectures that have so far resulted in extremely poor contacts.[13,18,24,32,54,55]

**Conclusion**

We have demonstrated the synthesis of a new type of GNR that can be *locally* switched from a semiconducting state to a metallic state that is well-described by a quasi-1D zigzag ladder Hamiltonian. The metallic state is unusual in that it is characterized by *two* broadly dispersive bands that cross $E_F$, making it a candidate two-channel conductor. Our ability to *write* atomically-precise metal-semiconductor heterojunctions into single GNRs presents new opportunities for low-barrier *p*-type contacts and future all-carbon nanoelectronic device architectures.

**Acknowledgement**

This work was primarily funded by the US Department of Energy (DOE), Office of Science, Basic Energy Sciences (BES), Materials Sciences and Engineering Division, under contract DE-AC02-05-CH11231 (M.F.C., F.R.F, S.G.L.) (Nanomachine program KC1203) (on-surface synthesis, STM imaging, DFT calculations, theoretical analyses) and contract DE-SC0023105 (F.R.F.) (molecular design). Support was also provided by the Office of Naval Research under awards N00014-19-1-2596 (M.F.C.) (image analysis) and N00014-24-1-2134 (F.R.F.) (molecular synthesis), and by the National Science Foundation under Grant No. CHE-2204252 (M.F.C.) (STM spectroscopy) and Grant No. DMR-2325410 (S.G.L.) (Wannier analysis and simulations). This research used resources of the National Energy Research Scientific Computing Center (NERSC), a U.S. Department of Energy Office of Science User Facility operated under Contract No. DE-AC02-05CH11231. Z.W. and B.Q. acknowledge support from Kavli ENSI Graduate Student Fellowships. R.Z. was supported by NAWA—Polish National Agency for Academic Exchange, grant number BPN/BEK/2023/1/00134. F.R.F. acknowledges support by the Heising-Simons Faculty Fellows

Program at UC Berkeley (development of synthetic facilities). We thank Prof. Eli Yablonovitch for valuable discussions. We thank Dr. Hasan Çelik and the UC Berkeley NMR facility in the College of Chemistry (CoC-NMR) for assistance with spectroscopic characterization. Instruments in the CoC-NMR are supported in part by National Institutes of Health (NIH) award no. S10OD024998. We thank Dr. Zhongrui Zhou at the UC Berkeley QB3 Chemistry Mass Spectrometry Facility for assistance with mass spectroscopic characterization.

**Author Contributions**

Z.W., K.S., W.T. and B.Q. initiated and conceived the research, K.S., B.Q., and F.R.F. designed the molecular precursors, K.S. synthesized, and characterized the molecular precursors, Z.W., R.Z., Y.H., X.L., P.H.J. and M.F.C. performed on-surface synthesis and STM characterization, Z.W., R.Z., H.P. and M.F.C. analysed the STM data, W.T. and S.G.L. developed the theory and performed DFT calculations, W.T., B.Q, S.G.L. and F.R.F. assisted with data interpretation, S.G.L., F.R.F., and M.F.C. acquired the funding and supervised the project. All authors contributed to scientific discussions and writing of the manuscript.

**Competing Interest**

The authors declare no competing interest.

## Methods

**Precursor Synthesis.** Full details of the synthesis and characterization of molecular precursor **1** are given in the Supplementary Information.

**Sample Preparation.** Atomically-clean Au(111) substrates were prepared using alternating $Ar^+$ sputtering and annealing cycles under UHV conditions. Sub-monolayer coverages of monomer **1** were prepared via UHV deposition using a homebuilt Knudsen cell evaporator at a crucible temperature of 130 °C for 10 min with the substrate held at 24 °C. Polymerization of **1** was achieved by ramping the substrate temperature by ~5 °C min$^{-1}$ to $T$ = 180 °C and holding this temperature for 20 min. Cyclodehydrogenation was induced by ramping the substrate temperature by ~15 °C min$^{-1}$ to $T$ = 300 °C and holding this temperature for 20 min.

**STM Measurements.** The molecule-decorated substrate was characterized at each stage of growth using a commercial Createc low-temperature STM operated at $T$ = 4.7 K with a chemically etched tungsten tip. STS measurements were conducted using a lock-in amplifier with a modulation frequency $f$ = 533 Hz and a modulation voltage amplitude of $V_{ac}$ = 4 mV. d$I$/d$V$ maps were recorded in constant-height mode with CO-functionalized tips and a modulation voltage amplitude of $V_{ac}$ = 30 mV. BRSTM images were obtained by mapping the d$I$/d$V$ signal at zero-bias and constant-height using a CO-functionalized tip. The parameters for STS peaks were determined by fitting them with Lorentzian line shapes. The average properties for each feature were extracted from ~20 spectra obtained from six different GNRs using different STM tips. STM images were processed using the WSxM software[56] package.

**STM Tip Manipulation.** Tip-induced dehydrogenation was performed by placing the STM tip above the site of a 5-membered ring at an initial tunnel current of 10–200 pA and $V_S$ = ±0.2 V. The current feedback loop was then disabled and the sample voltage was ramped from 0 to ±2.5 V and then back to 0 V while recording the tunnel current.

**Calculations.** First-principles DFT-LDA calculations were performed using the Quantum ESPRESSO package.[57,58] Norm-conserving (NC) pseudopotentials were employed[59,60] with an energy cutoff of 90 Ry and Gaussian smearing of 0.002 Ry. A vacuum layer of 10 Å was applied along all non-periodic directions to ensure the accuracy of the results. Chemical structures were fully relaxed until the forces on all atoms were smaller than 0.01 eV $Å^{-1}$. Band structure calculations were performed using 40 ***k***-points along the periodic direction. The Wannier90 package[61] was used for performing Wannierization from DFT-LDA results, and the PythTB package[62] was used to obtain tight-binding parameters.

**Data Availability**

All data reported in the manuscript is provided as part of source data files. The Supplementary Information contains Wannier analysis of 7-iGNRs (Note S1); materials and methods (Note S2); detailed synthetic procedures and characterization of precursor **1** (Scheme S1); NMR characterization of intermediates and products (Figure S1–S4); and references.

**Code Availability**

The DFT code and pseudopotentials can be downloaded from the Quantum Espresso website.[57,63] We used version 6.7 for the DFT-LDA calculations performed here. The Wannier90 code can be downloaded from the Wannier90 website.[64] The tight-binding simulation code can be downloaded from the PythTB website.[62] We used version 1.8.0 for the tight-binding fitting performed here.

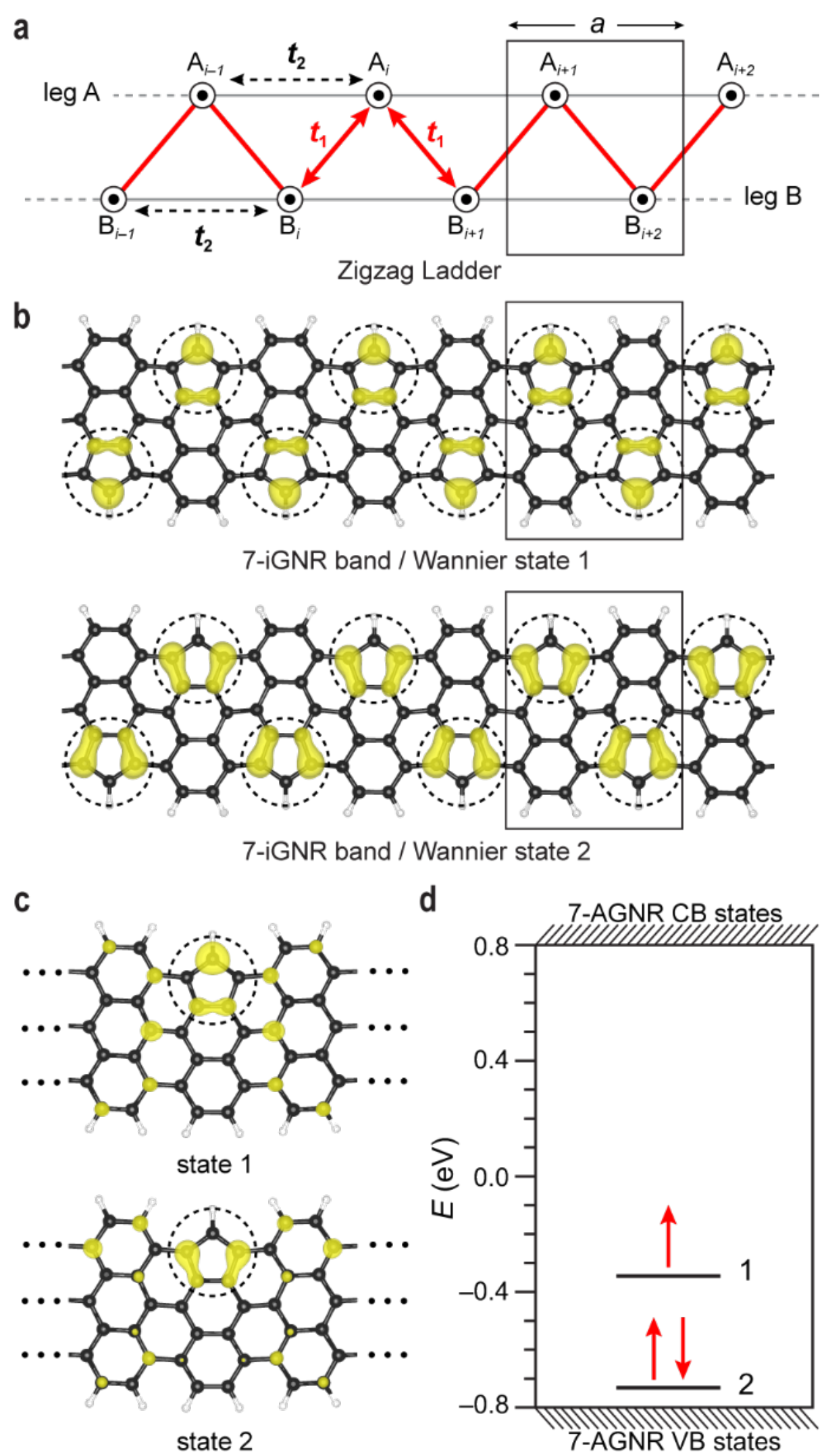

**Figure 1 | Expanded SSH model on a zigzag ladder. a,** Tight-binding (TB) model for electrons on a zigzag ladder. Each unit cell (boxed region with lattice constant = $a$) contains two onsite orbitals located at sites $A_i$ and $B_i$ belonging to legs A and B, respectively, with $i$ denoting the site index. Red arrows describe NN hopping $t_1$, black dashed arrows describe NNN hopping $t_2$. **b,** Sketch of 7-iGNR structure including schematic representation of the two independent Wannier function networks (Wannier state 1 (top), Wannier state 2 (bottom)) that reside primarily on the 5-membered rings and result in an effective zigzag ladder. **c,** DFT-calculated wavefunction of the two defect states (*state* 1 and *state* 2) arising upon introduction of a single 5-membered ring. **d,** DFT-calculated energy level diagram of the two defect states that appear in the bandgap of a 7-AGNR due to introduction of a single 5-membered ring.

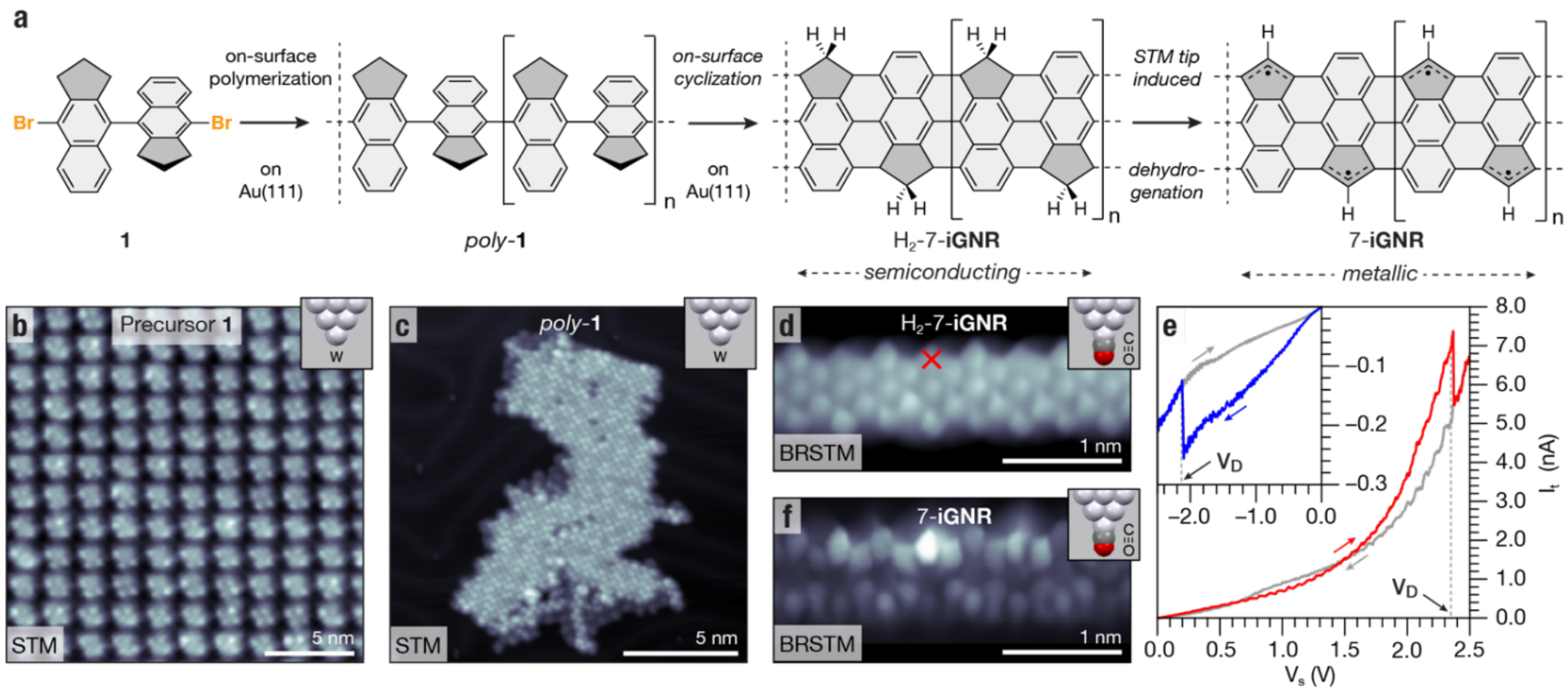

**Figure 2 | On-surface synthesis of 7-iGNRs. a,** Schematic representation of the stepwise on-surface synthesis of 7-iGNRs from **1** on Au(111). STM tip-induced dehydrogenation of $H_2$-7-iGNR enables conversion to 7-iGNR. **b,** STM topographic image of self-assembled island of molecular precursor **1** thermally evaporated onto Au(111) (sample bias $V_s$ = –500 mV, tunnel current $I_t$ = 50 pA). **c,** STM topographic image of polymerized *poly*-**1** on Au(111) after annealing at $T$ = 180 °C for 20 min ($V_s$ = –800 mV, $I_t$ = 50 pA). **d,** Constant-height bond-resolved STM (BRSTM) image of a segment of $H_2$-7-iGNR after additional annealing at $T$ = 300 °C for 20 min ($V_s$ = 0 mV, $V_{ac}$ = 30 mV, $f$ = 533 Hz, CO-functionalized tip). **e,** $I_t$–$V_s$ plot showing hysteric voltage sweep over range 0.0 V ≤ $V_s$ ≤ +2.5 V (inset: sweep over –2.5 V ≤ $V_s$ ≤ 0.0 V). An irreversible discontinuity in the tunnel current at $V_s$ > +2.0 V ($V_s$ < –2.0 V) represents the threshold voltage for STM tip-induced dehydrogenation $V_D$. **f,** Constant-height BRSTM image of the same GNR segment as in **d** following the application of a STM tip pulse ($V_s$ = –2.5 V for 300 ms) at the position marked by a red cross ($V_s$ = 0 mV, $V_{ac}$ = 30 mV, $f$ = 533 Hz, CO-functionalized tip). All STM experiments were performed at $T$ = 4.7 K.

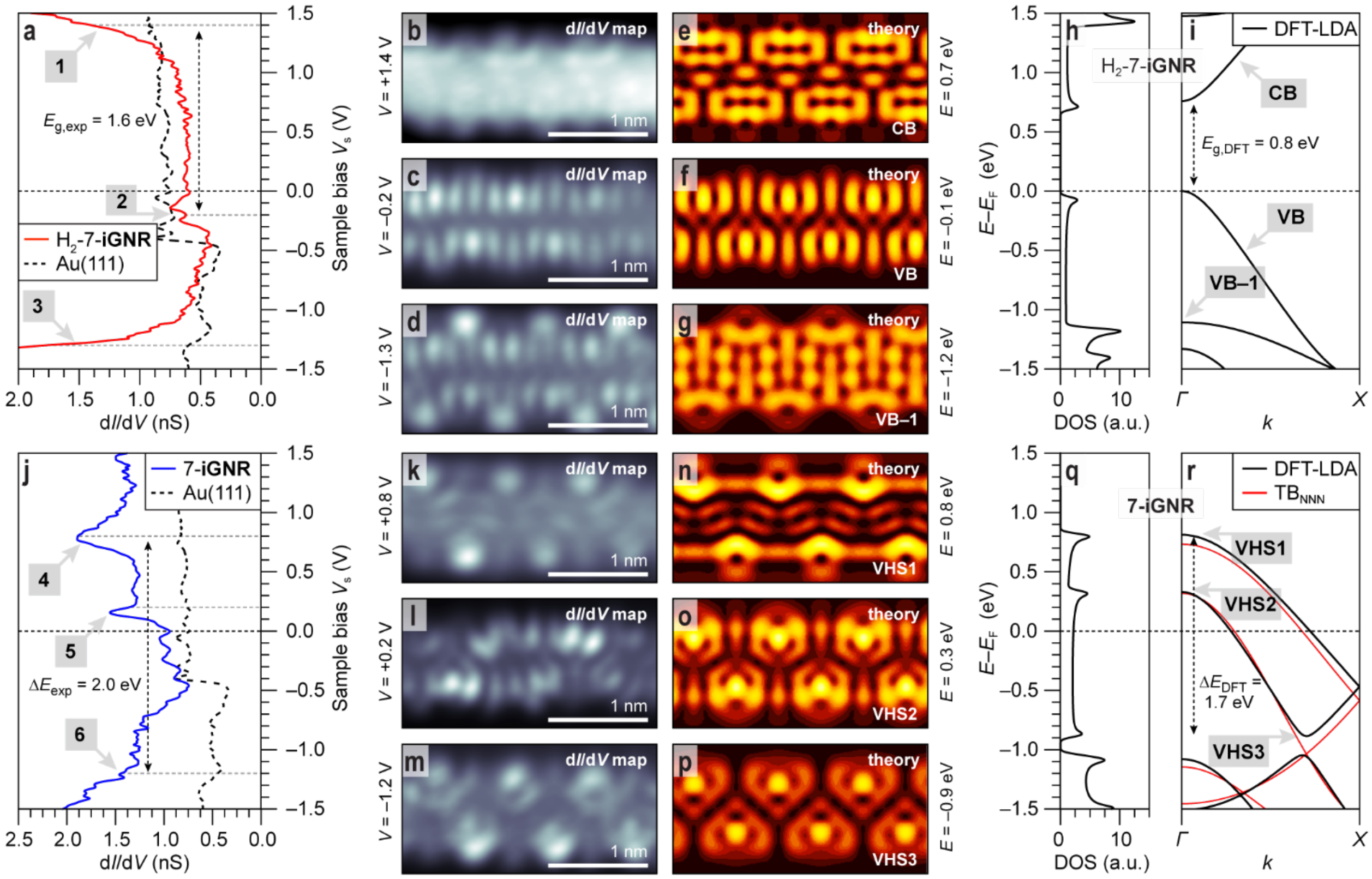

**Figure 3 | Electronic structure of $H_2$-7-iGNRs and 7-iGNRs. a,** d$I$/d$V$ point spectrum of an $H_2$-7-iGNR on Au(111) recorded at the position of a 5-membered ring ($V_{ac}$ = 30 mV, $f$ = 533 Hz; setpoint: $V_s$ = 1.5 V, $I_t$ = 2 nA, CO-functionalized tip). **b–d,** Constant-height d$I$/d$V$ maps recorded at sample voltage biases (**b**) $V_s$ = +1.4 V (**c**) $V_s$ = –0.2 V, and (**d**) $V_s$ = –1.3 V ($V_{ac}$ = 30 mV, $f$ = 533 Hz, CO-functionalized tip). **e–g,** Calculated DFT-LDA LDOS maps for a freestanding $H_2$-7-iGNR at energies corresponding to (**e**) the bottom of the CB, (**f**) the top of the VB, and (**g**) the top of the VB–1. **h,** Calculated DFT-LDA DOS for a freestanding $H_2$-7-iGNR (spectrum broadened by 27 meV Gaussian). **i,** DFT-LDA band structure of freestanding $H_2$-7-iGNR. **j,** d$I$/d$V$ point spectrum of a 7-iGNR on Au(111) recorded at the position of a 5-membered ring ($V_{ac}$ = 30 mV, $f$ = 533 Hz; setpoint: $V_s$ = 1.5 V, $I_t$ = 2 nA, CO-functionalized tip). **k–m,** Constant-height d$I$/d$V$ maps recorded at sample voltage biases (**k**) $V_s$ = +0.8 V, (**l**) $V_s$ = +0.2 V, and (**m**) $V_s$ = –1.2 V ($V_{ac}$ = 30 mV, $f$ = 533 Hz, CO-functionalized tip). **n–p,** Calculated DFT-LDA LDOS maps for a freestanding 7-iGNR at energies corresponding to (**n**) VHS1, (**o**) VHS2, and (**p**) VHS3. **q,** Calculated DFT-LDA DOS for a freestanding 7-iGNR (spectrum broadened by 27 meV Gaussian). **r,** DFT-LDA band structure (black) for a freestanding 7-iGNR. Band structure derived from the NNN tight-binding Hamiltonian in the Wannier basis is shown in red. All STM experiments were performed at $T$ = 4.7 K. Theoretical LDOS maps are sampled at a height of 3.5 Å above the atomic plane of the GNR.

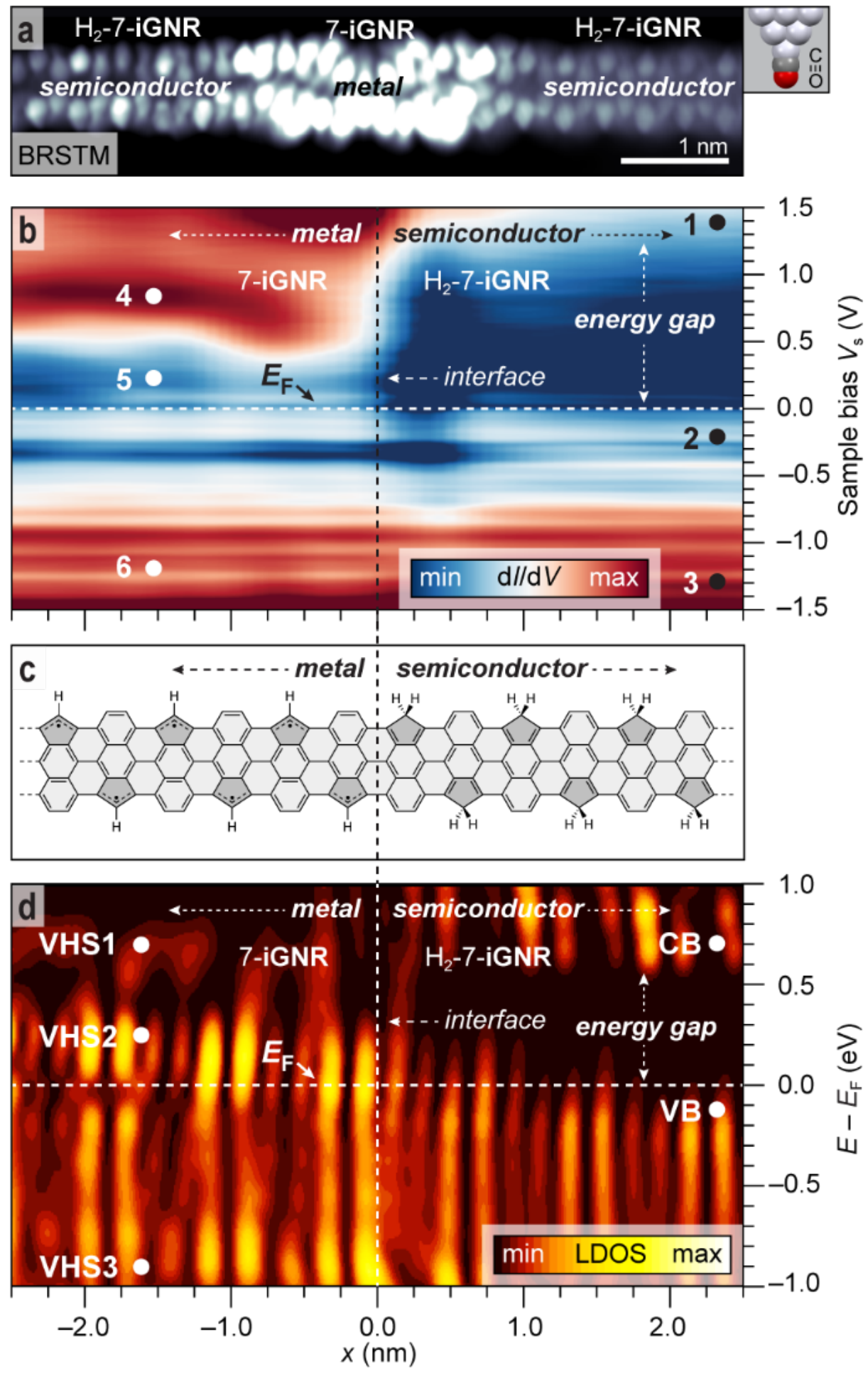

**Figure 4 | STM tip-induced metal-semiconductor heterojunctions. a,** Constant-height BRSTM image of a semiconducting $H_2$-7-iGNR featuring a metallic 7-iGNR segment (3 unit cells) embedded in the center of the ribbon ($V_s$ = 0 mV, $V_{ac}$ = 30 mV, $f$ = 533 Hz, CO-functionalized tip). **b,** STS d$I$/d$V$ ($V_s$, $x$) acquired across the metal-semiconductor interface of a single-GNR 7-iGNR/$H_2$-7-iGNR heterojunction. The interface is located at $x$ = 0 nm with an uncertainty of ±2 Å. The main d$I$/d$V$ features of pristine 7-iGNR and $H_2$-7-iGNR are marked for reference (numbered dots). **c,** Representative structure of 7-iGNR/$H_2$-7-iGNR single-ribbon metal-semiconductor heterojunction with interface at $x$ = 0. **d,** LDOS calculated along linescan traversing the metal-semiconductor interface of the freestanding 7-iGNR/$H_2$-7-iGNR heterojunction shown in (**c**). Calculated DFT-LDA DOS main features for freestanding 7-iGNR and $H_2$-7-iGNR are marked for reference (labelled dots).

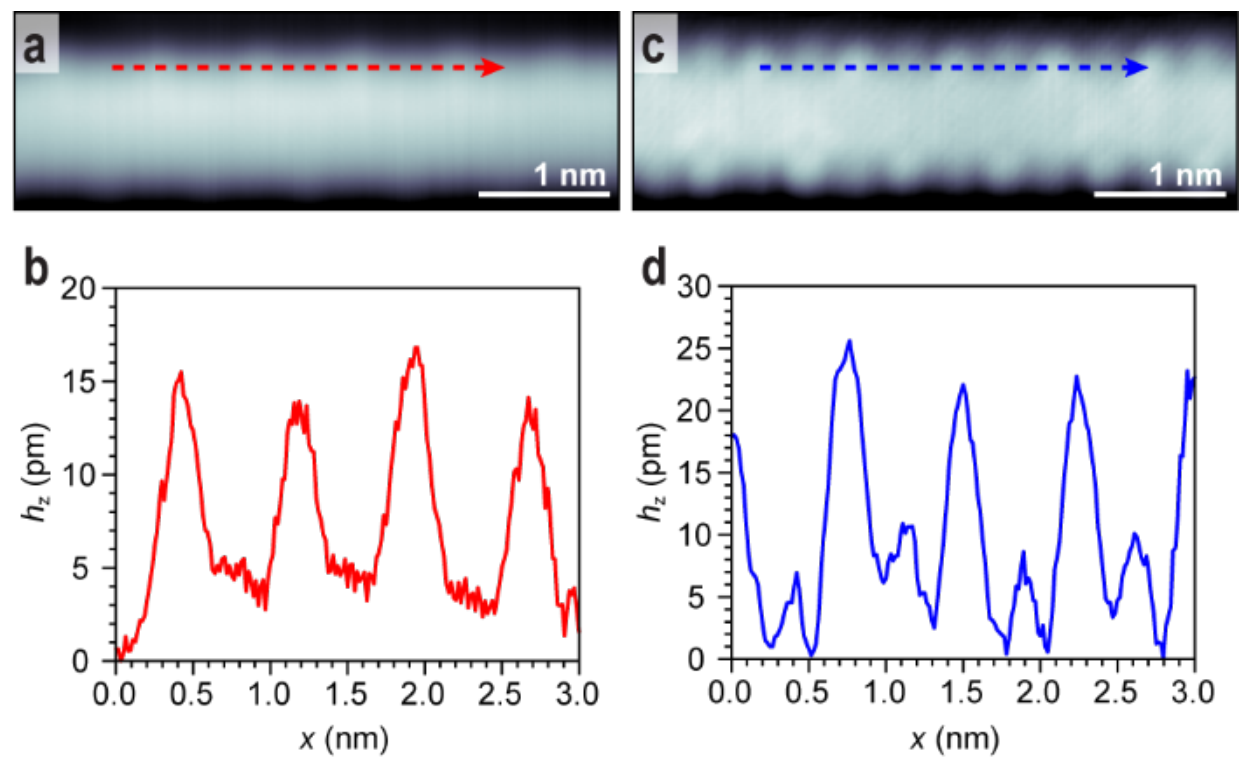

**Extended Data Figure 1 | Height variation for STM topography of $H_2$-7-iGNR and 7-iGNR. a,** Topographic STM image of $H_2$-7-iGNR ($V_s$ = +700 mV, $I_t$ = 50 pA, CO-functionalized tip). **b,** Height variation along line scan for $H_2$-7-iGNR (red dashed arrow in (**a**)). **c,** Topographic STM image of 7-iGNR ($V_s$ = −500 mV, $I_t$ = 50 pA, CO-functionalized tip). **d,** Height variation along line scan for 7-iGNR (blue dashed arrow in (**c**)). All STM experiments performed at $T$ = 4.7 K.

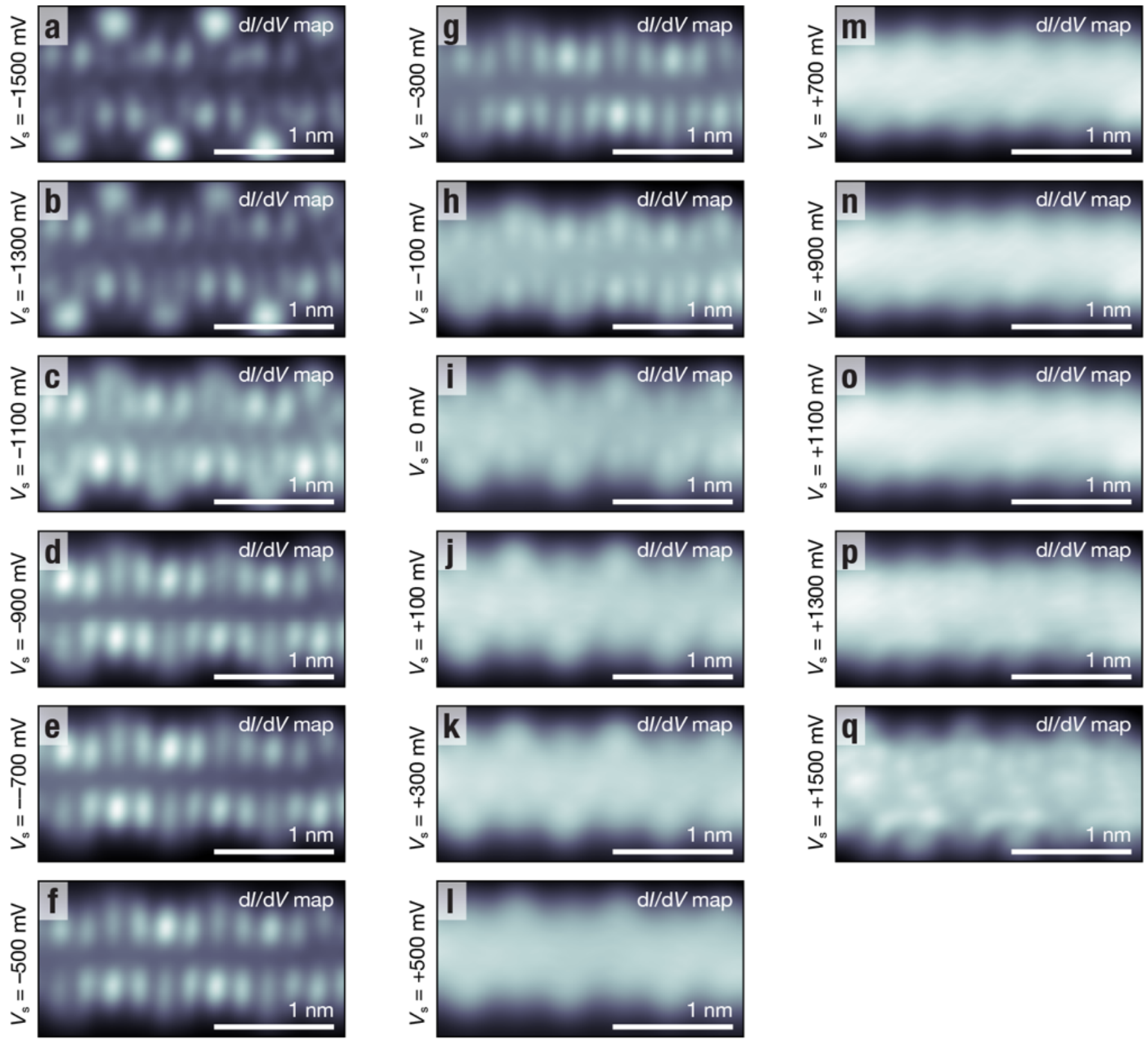

**Extended Data Figure 2 | d*I*/d*V* maps of $H_2$-7-iGNR. a–q,** Constant-height d*I*/d*V* maps recorded at the indicated sample biases ($V_{ac}$ = 30 mV, *f* = 533 Hz, CO-functionalized tip). All STM experiments performed at *T* = 4.7 K.

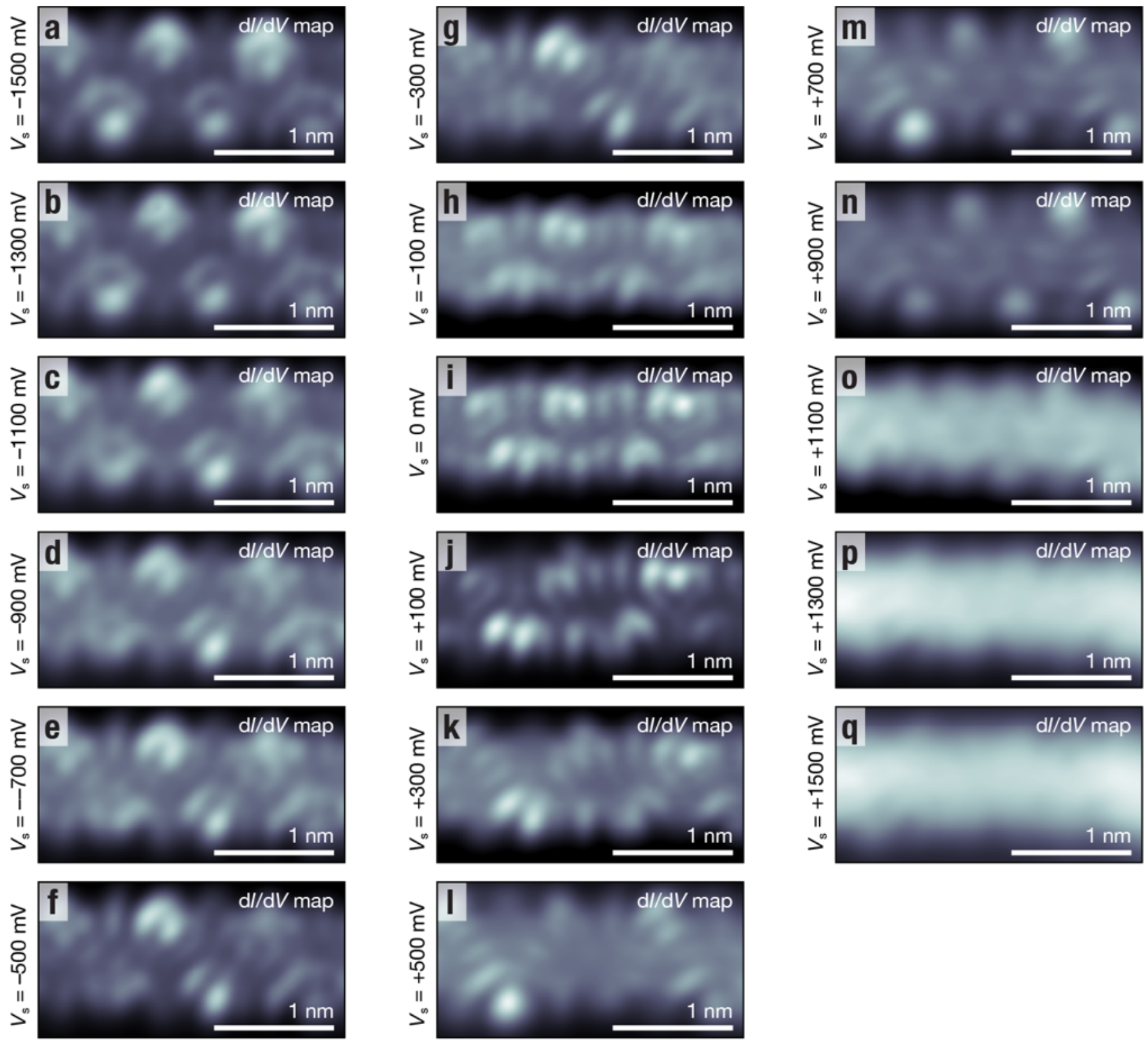

**Extended Data Figure 3 | d*I*/d*V* maps of 7-iGNR. a–q,** Constant-height d*I*/d*V* maps recorded at the indicated sample biases ($V_{ac}$ = 30 mV, $f$ = 533 Hz, CO-functionalized tip). All STM experiments performed at $T$ = 4.7 K.

| | $\varepsilon – E_F$ | $t_1$ | $t_2$ |
|---|---|---|---|
| Wannier network *1* | –0.474 | 0.056 | 0.547 |
| Wannier network *2* | –1.014 | 0.300 | 0.366 |

**Extended Data Table 1 |** Parameters for the Wannier tight-binding fitting using Eq. (2). All energies are in eV.